\documentclass[runningheads]{llncs}
\usepackage[T1]{fontenc}
\usepackage{graphicx} 
\usepackage{macros}
\usepackage{caption}
\usepackage{tikz,pgfplots}
\usepackage[left=3.5cm,right=3.5cm]{geometry}
\usepackage{makecell,graphicx}
\usepackage{amssymb,comment,xspace}
\usepackage[OT1]{fontenc}

\usepackage{newtxtext}        
\usepackage[varvw]{newtxmath}      
\usepackage{circuitikz}
\usepackage{thmtools}
\usepackage{thm-restate}
\newcommand{\explosions}[1]{|\mathord{\divideontimes}(#1)|}
\newcommand{\ruleone}{{\mbox{\textsf{Rule 1 (fat)}}}\xspace}
\newcommand{\ruletwo}{{\mbox{\textsf{Rule 2 (delete)}}}\xspace}
\newcommand{\rulethree}{{\mbox{\textsf{Rule 3 (cheap)}}}\xspace}
\newcommand{\rulefour}{{\mbox{\textsf{Rule 4 (medium)}}}\xspace}
\newcommand{\rulefive}{{\mbox{\textsf{Rule 5 (costly)}}}\xspace}
\newcommand{\rulesix}{{\mbox{\textsf{Rule 6 (many)}}}\xspace}

\crefname{appsec}{Appendix}{Appendices}   
\crefalias{appsec}{section} 

\let\oldproof\proof
\let\endoldproof\endproof

\renewenvironment{proof}
  {\oldproof}
  {\qed\endoldproof}
\begin{document}
\title{Integrality Gap of 3.4 for Santa Claus: A Topology-based Analysis}

\author{Kimon Boehmer\inst{1,2}\orcidID{0009-0002-5831-6158 }
\and
Chien-Chung Huang\inst{1}\orcidID{0000-0001-5223-0770}\\
\email{\{kimon.bohmer, chien-chung.huang\}@ens.fr}
}
\authorrunning{Boehmer and Huang}
%
\institute{DI ENS, École normale supérieure, Université PSL, Paris, France \and
LIP, ENS Lyon, France}
\maketitle             
\begin{abstract}
In the \textsc{Max-Min Allocation} problem, a set of indivisible items are to be assigned to a set of players and the goal is to maximize the minimum value among all players. A much studied special case, the \textsc{Santa Claus} 
    problem, assumes that an item has a fixed utility but 
    a player is only interested in a subset of items. 
    
    The main technique for designing approximation algorithms 
    for \textsc{Santa Claus} is the configuration LP, 
    introduced by the Bansal and Sviridenko. A long line of research has progressively improved the upper bound of the integrality gap, mostly relying on  the combinatorial arguments of Haxell 
    for finding perfect matchings in certain hypergraphs. 
    Remarkably, a recent paper of Haxell and Szabó (SODA '23) introduced a novel topology-based analysis, by which 
    they are able to obtain the current best upper bound of 
    $3.534$.
    
    In this work, we adapt the approach of Haxell and Szab\'o 
    to analyze the integrality gap. The main new ingredient in our adaptation is that we stop the process of ``edge explosions'' in a graph used by Haxell and Szab\'o when it becomes 
    too costly. This early termination of explosions, combined with a new lower bound on a topological connectivity parameter based on a theorem of Aharoni, Berger, and 
    Meshulam, allows us to improve the integrality gap to 3.4. Our analysis has the additional advantage of being simpler. 
\end{abstract}

\newpage
\section{Introduction}
In the \textsc{Max-Min Allocation} problem, 
we are given
an item set $I$ and a player set $P$. 
Each player $p$ has a value function
$\nu_p: I \rightarrow \RR^+$ over the items. 
An allocation $A: P \rightarrow 2^{I}$ 
assigns items to players so that each 
item is given to at most one player. 
The objective is to find an allocation $A$ 
that maximizes $\min_{p \in P}\sum_{i \in A(p)}\nu_p(i)$. Put differently, we seek 
a \emph{fair} allocation. 

A special case, the \emph{restricted assignment}, has attracted much attention. Here each item has a fixed value  but 
only a subset of players are interested in it. More formally, let $a: I \rightarrow \RR^+$ be a value function. 
Each player $p \in P$ is interested in a set $R_p \subseteq I$. They have value $\nu_p(i) = 0$ for an item $i \in I - R_p$ 
and $\nu_p(i)=a(i)$ for an item $i \in R_p$.
In this setting, which often goes by the name of \textsc{Santa Claus}, the protagonist needs to ensure that each child (player) receives desired gifts (items) as valuable as possible. 

The main tool to design approximation algorithms for \textsc{Santa Claus} is the  \emph{configuration LP}, introduced by Bansal and Sviridenko~\cite{bansal2006santa}, which we briefly summarize below.   
Let $T \geq 0$. We define a family 
$\mathcal{C}(T,p):=\{C \subseteq R_p \mid \sum_{i \in C}a(i) \geq T\}$ of \emph{configurations} for player $p$. In other words, each configuration $C \in \mathcal{C}(T,p)$ has value at least $T$ for player $p$. For every player $p$, we associate a variable $x_{p,C}$ with each configuration  $C \in \mathcal{C}(T,p)$. 

\begin{align*}
    \sum_{C \in \mathcal{C}(T,p)} x_{p,C} &\geq 1 & \forall p \in P \nonumber\\
    \sum_{p \in P}\sum_{C \in \mathcal{C}(T,p): i \in C} x_{p,C} &\leq 1 & \forall i \in I\nonumber\\
    x_{p,C} &\geq 0 & \forall p \in P, \forall C \in \mathcal{C}(T,p)\nonumber\tag{CLP}\label{lp:clp}
\end{align*}
Assuming that this LP minimizes the constant zero vector, we can write the dual of \eqref{lp:clp} as
\begin{align*}
    \max \sum_{p \in P} y_p &- \sum_{i \in I} z_i &  \text{s.t.}\nonumber\\
    \sum_{i \in C} z_i &\geq y_p &\forall p \in P, \forall C \in \mathcal{C}(T,p)\nonumber\\
    y_p,z_i &\geq 0 & \forall i \in I,p \in P\nonumber \tag{Dual CLP}\label{lp:dclp}
\end{align*}

Let $T^*$ be the maximum of $T$ for which \eqref{lp:clp} is feasible. Although \eqref{lp:clp} has exponentially many variables, a $(1-\varepsilon)$-approximate solution can be found in polynomial time for any desired fixed accuracy $\varepsilon>0$~\cite{bansal2006santa}. 
  Let {\sf OPT} denote the optimal objective value in the given instance. Then clearly ${ \sf OPT} \leq T^*$. The integrality gap of \eqref{lp:clp} is defined as $\frac{T^*}{{\sf OPT}}$. To show the integrality gap is upper-bounded by $\alpha^{-1}$, it suffices to prove the existance of an allocation $A$ so that
$\min_{p \in P}\sum_{i \in A(p)}a(i) \geq \alpha T^*$.
\paragraph{The story so far and our contribution.}
\begin{table}[h]
\centering
\caption{Known results on the upper bound of the integrality gap of \eqref{lp:clp}.}
\label{tab:summary_streaming}
\renewcommand{\arraystretch}{1.2}

\begin{tabular}{ m{3.8cm} m{2.2cm}  m{6.5cm}}
\hline
 Algorithm & Integrality Gap & Details  \\ \hline
 Bansal and Sviridenko~\cite{bansal2006santa} & $\OO(\frac{\log \log |P|}{\log \log \log |P|})$  & constructive in polynomial time  \\ \hline
Feige~\cite{feige2008allocations}  & $O(1)$  &   non-constructive; could be made constructive  in polynomial time using the technique of Haeupler, Saha and Srinivasan~\cite{haeupler2011new} \\ \hline
Asadpour, Feige, Saberi~\cite{asadpour2012santa} & 4 & non-constructive; 
could be made constructive in quasi-polynomial time using the technique of Polacek and Svensson~\cite{polavcek2015quasi} \\ \hline
Annamalai, Kalaitzis, Svensson~\cite{annamalai2017combinatorial}& 12.33 & constructive in polynomial time \\\hline
Davies, Rothvoss, Zhang~\cite{davies2020tale}, Cheng and Mao~\cite{cheng2018restricted}& 6 & constructive in polynomial time \\\hline
Cheng and Mao~\cite{cheng2019restricted}& 4 & constructive in polynomial time \\\hline
Jansen and Rohwedder~\cite{jansen2020note}, Cheng and Mao~\cite{cheng2018integrality} & 3.833 & non-constructive \\\hline
Cheng and Mao~\cite{cheng2019restricted} & 3.808 & non-constructive \\\hline
Haxell and Szab\'{o}~\cite{haxell2025improved} & 3.534 & non-constructive \\\hline 
\textbf{Present Work} & \textbf{3.4} & \textbf{non-constructive} \\\hline
\end{tabular}
\end{table}
An example showing that the integrality gap of \eqref{lp:clp} is at least $2$ is given by Chan, Tang and Wu~\cite{chan20181}.
Bansal and Sviridenko~\cite{bansal2006santa} established that the integrality gap of \eqref{lp:clp} is at most $\OO(\frac{\log \log |P|}{\log \log \log |P|})$ and Feige~\cite{feige2008allocations} first gave a constant bound. 
A sequence of follow-up works have progressively improved the upper bound further, see Table~\ref{tab:summary_streaming} for a summary. Notice that 
some of these results are \emph{non-construtive}, i.e. the proof on the gap is existential and does not immediately lead to a polynomial time algorithm to find an actual allocation achieving the approximation ratio promised by the integrality gap. However, by solving \eqref{lp:clp} using the approach of~\cite{bansal2006santa}, these non-constructive proofs still imply that one can estimate the \emph{value} of the optimal solution in polynomial time. 

The current best bound of $3.534$ is attained by Haxell and Szabó~\cite{haxell2025improved}. 
Their bound is even more remarkable as it is achieved by a novel topological argument, unlike most previous works~\cite{asadpour2012santa,annamalai2017combinatorial,cheng2019restricted,cheng2018integrality,cheng2018restricted,jansen2020note} that are based on a combinatorial augmenting tree argument of Haxell~\cite{haxell1995condition} for finding perfect matchings in certain hypergraphs. 

In this work, we will 
develop further on the 
topology-based approach 
of Haxell and Szab\'{o} to 
analyze the integrality 
gap. Our main contribution is the following. 

\begin{theorem} \label{th:main}
    The integrality gap of \eqref{lp:clp} is bounded by $3.4$.
    \label{thm:main}
\end{theorem}    
In addition to the improvement in the constant, our analysis is cleaner and simpler, as we will explain shortly.
Note that our approach, similar to that of Haxell and Szab\'{o}, does not guarantee a polynomial-time algorithm to construct the corresponding allocation. 
\paragraph{Our technique.}
We sketch the high-level ideas of the topology-based approach of Haxell and Szab\'o and how we modify it in order to obtain our improvments. The more precise definitions and details will be given in Section~2. 

Let $\alpha \in (0,1)$ and suppose that the goal is to establish that the integrality gap is $\alpha^{-1}$. Assume that a configuration threshold $T>0$ is given for which \eqref{lp:clp} is feasible.  
Let $H$ be a $|P|$-partite graph, where each player $p\in P$ is associated with a set $V^p$ of vertices. A vertex $v^p \in V^P$ corresponds to a subset of items in $R_p$ and these items have the total value of at least $\alpha T$. In $H$, two vertices $v^p \in V^P$ and $u^q \in V^q$ of two different players $p,q \in P$ are connected by an edge if the two sets of items to which they correspond intersect. 
By this definition of $H$, if we can find an \emph{independent transversal}, which is an independent set containing one vertex in $V^p$ of each player $p \in P$, then this transversal defines an allocation, where each player attains the value of $\alpha T$, thus guaranteeing the desired integrality gap of $\alpha^{-1}$. 

A sufficient condition for the existence of such an independent transversal is given by a Hall-marriage-style theorem (Theorem~\ref{th:hall-general}) first observed by Aharoni~\cite{aharoni2005eigenvalues}. 
It states that an independent transversal exists if, for every subset $Q \subseteq P$ of players, for the the subgraph $H|_{Q}$ of $H$ that is induced by $Q$, a certain topological parameter $\eta(H|_{Q})$ 
(which will be formally defined in Section~2) is at least $|Q|$. 

To lower bound this parameter $\eta(H|_{Q})$, Haxell and Szab\'o use a theorem of Meshulam~\cite{meshulam2003domination} (Theorem~\ref{th:meshulam}), which essentially lower-bounds $\eta(H|_{Q})$ by the number of possible \emph{explosions} that can be performed in $H|_Q$.
An explosion consists of picking a remaining edge $(v^p, u^q)$ in $H|_{Q}$ and removing $v^p$, $u^q$ and all their neighbors. 
To lower-bound the number of explosions, Haxell and Szab\'o  construct a pair of feasible dual variables $(y,z)$ in \eqref{lp:dclp}. 

We now explain where our approach diverges from theirs. Our critical observation is that the explosions become more and more ``expensive''. This observation suggests that we stop the explosion process early and try to build the dual variables $(y,z)$ on the spot. The early termination deprives us some of the properties that are guaranteed in the original analysis of \cite{haxell2025improved}, however, thanks to a theorem of 
Aharoni, Berger and Meshulam~\cite{aharoni2005eigenvalues}, 
we are still able to define the desired duals $y$ and $z$ to lower-bound the number of explosions.
Their theorem implies that, assuming that $H^*$ is the final graph when the explosion process stops, $\eta(H^*)$ can be lower-bounded by the optimal fractional vertex cover of a graph closely related to $H^*$. 
In this way, our dual variables of \eqref{lp:dclp} depend not only on the value of the items, but also on the occurrences of the items in $H^*$.

Our analysis is simpler in that we use less ``snapshots''. In~\cite{haxell2025improved}, 
there are four checkpoints in the whole process of explosions. At each checkpoint, a lower bound on the number of explosions of various types is established by demonstrating 
a feasible dual pair $(y,z)$. The final ratio of $ \approx3.53$ is then obtained by a convex combination of the four different lower bounds.~\footnote{To be precise, in~\cite{haxell2025improved}, they also give a simpler analysis of integrality gap 4 using just a single snapshot.} 

In comparison, we only use two snapshots. In fact, by using just a single snapshot, we can already give 
an integrality gap of $3.5$. This single-snapshot version of our analysis is simpler, so we will 
first present it in \Cref{sec:simpler}. To improve the integrality gap down to $3.4$, we first use a snapshot whose analysis is already done in~\cite{haxell2025improved} (Lemma~\ref{lem:lp-application-1}). The second snapshot is a finer analysis of the single snapshot in Section~3. This finer analysis uses the additional idea of artificially modifying the value function of the items so as to narrow the gap between the upper lower bound of the ``cover'' of exploded vertices, on which our analysis in \Cref{sec:simpler} is based.
Our gap of 3.4 is proved using a convex combination of these two snapshots.
\paragraph{Related work.}
It is worth mentioning that for the more general \textsc{Max-Min Allocation} problem, the integrality gap is shown to be much larger,  namely, $\Omega(\sqrt{|P|})$, by Bansal and Sviridenko~\cite{bansal2006santa}. 
 The current best approximation ratio is $|P|^{\varepsilon}$ for any $\varepsilon>0$, due to Chakrabarty, Chuzhoy and Khanna~\cite{chakrabarty2009allocating}. However, the best hardness of approximation is only $2$, established by Bezáková and Dani~\cite{bezakova2005allocating}.

For the special case of~\textsc{Santa Claus} where all players can receive all items, namely, $R_p=I$ for all $p \in P$, Woeginger~\cite{woeginger1997polynomial}, and Epstein and Sgall~\cite{epstein2004approximation} have designed approximation schemes. Boehmer~\cite{boehmer2025approximability} has shown that the integrality gap of the configuration LP is between $1+\frac{1}{72}$ and $1.5$ in that case.

Another studied special case of \textsc{Santa claus} is the $(1,\varepsilon)$-restricted assignment, where each item has value $1$ or $\varepsilon \in (0,1)$. Chan, Tang and Wu~\cite{chan20181} show that in this case, the integrality gap of \eqref{lp:clp} is at most 3. Haxell and Szab\'{o}~\cite{haxell2025improved} show that the gap is strictly less than $3$ in most cases.
\paragraph{Roadmap.}
In \Cref{sec:pre}, we introduce notation and definitions, and make some preliminary observations.
In \Cref{sec:simpler}, we present our main proof idea and give an analysis that bounds the integrality gap by $3.5$.
In \Cref{sec:fat}, we show how to improve this bound to $3.4$, proving our main result.
Some observations and proofs are deferred to the Appendix.

\section{Preliminaries}\label{sec:pre}
\paragraph{Short-hand notation.} 
Given a function $x: S' \rightarrow \RR^{+}$ 
and a subset $S \subseteq S'$, we write $x(S):=\sum_{s \in S} x(s)$.
Analogously, for a vector $x \in \RR^{S'}$, we define $x(S):=\sum_{s \in S} x_s$.
We write $S-s$ to denote $S-\{s\}$.

\paragraph{Graph notation.} 
Let $G=(V,E)$ be a graph. When we specifically refer to its vertex set $V$, we write $V(G)$. Similarly, its edge set will be written as $E(G)$. An edge in $E$ connecting two vertices $u$ and $v$ will be written as  $(u,v)$, but very often 
we will treat $(u,v)$ as a set of two vertices. 
The \emph{line graph} of $G$ contains a vertex for each edge of $G$, and two vertices are connected if the corresponding edges in $G$ share an endpoint.
Formally, the line graph of $G$ is $L(G)=(E,\{(e,e') \in E^2 \mid |e \cap e'|=1\})$.
Given a set $S \subseteq V$, the \emph{neighbourhood} of $S$ is defined as $N[S]:=S \cup \{u \in V \mid \exists v \in S: (v,u) \in E\}$.
The subgraph $G[S]$ \emph{induced} by $S$ is defined as $(S,\{e \in E \mid e \subseteq S\})$.
We denote $\delta(v) = \{e \in E \mid v \in e\}$.

Let $e \in E$.
The operation $G-e$ of \emph{deleting} $e$ from $G$ results in the graph $(V,E-e)$.
The operation $G \divideontimes e$ of \emph{exploding} $e$ in $G$ results in the graph $G[V-N[e]]$.
In other words, in $G \divideontimes e$, all neighbours of the two endpoints of $e$ are removed.

An \emph{independent set} of $G$ is a set $S \subseteq V$ such that $G[S]$ has no edges.
A \emph{fractional vertex cover} of $G$ is a vector $z' \in [0,1]^V$ such that $z'(e) \geq 1$ for all $e \in E$.
The \emph{minimum fractional vertex cover} of $G$ is the fractional vertex cover $z'$ minimizing $z'(V)$. We denote its value by $\tau^*(G)$.
\paragraph{Independent transversals.} 
We follow the methodology of \cite{haxell2025improved} by reformulating the task of finding an $\alpha$-approximate solution to \eqref{lp:clp} as a graph problem.
\begin{definition}[$\alpha$-approximation allocation graph]
    Let $\mathcal{I}$ be an instance of our problem, and let $T$ be the maximum configuration threshold for which \eqref{lp:clp} is feasible.
    A set $v \subseteq R_p$ is called a \emph{minimal set} of player $p$ if $a(v) \geq \alpha T$, but $a(v-j)< \alpha T$ for all $j \in v$.
    We define the \emph{minimal vertices} of $p$ as 
    $V^p:=\{v^p \mid v \mbox{ is minimal set of $p$}\}$.  
    The \emph{$\alpha$-approximation allocation graph} $H_{\mathcal{I}}(\alpha)$ is a $|P|$-partite graph with vertex set $\bigcup_{p \in P}V^p$ and there is an edge between $v^p$ and $u^q$ if $p \neq q$ and $v \cap u \neq \emptyset$.

 \label{def:allocationGraph}   
\end{definition}
An \emph{independent transversal} of $H_{\mathcal{I}}(\alpha)$ is an independent set that contains exactly one minimal vertex $v^p \in V^p$ for each $p \in P$. Clearly, such an independent transversal defines an allocation which is an $\alpha$-approximation to the optimal value of \eqref{lp:clp}.
In the main part of the analysis, we will fix an instance $\mathcal{I}$ and a parameter $\alpha$, and then we define $H:=H_{\mathcal{I}}(\alpha)$ with which we will work. 
The \emph{restriction} of $H$ to a subset $Q \subseteq P$ of players is defined as $H|_{Q}:=H[\bigcup_{p \in Q} V^p]$.

\paragraph{The homological connectivity parameter $\eta$.}
A central ingredient in the analysis of Haxell and Szabó is the topological graph parameter $\eta(G)$.
Given a graph $G$, let $I(G)$ be the independence complex of $G$, i.e. the abstract simplicial complex whose faces are the independent sets of $G$.
Then, $\eta(G)$ is defined as the \emph{homological connectivity} of $I(G)$ plus $2$.
We refer to \cite{haxell2025improved} for intuition and a friendly introduction to this parameter.
In fact, readers who are unfamiliar with homological connectivity can treat $\eta$ in a black box manner, as 
all its required properties and tools will be summarized in the following discussion. 

Recall that our goal is to prove that $H$ contains an independent transversal.
The following theorem provides a Hall-like condition on the existence of independent 
transversals.
\begin{theorem}[Aharoni]\label{th:hall-general}
    If $\eta(H|_{Q})\geq |Q|$ for all $Q \subseteq P$, then $H$ contains an independent transversal.
\end{theorem}
Therefore, our main goal is to lower-bound $\eta$.
A common approach is to use a technique by Meshulam.
An edge $e \in E(G)$ is \emph{deletable} if $\eta(G-e) \leq \eta(G)$.
An edge $e$ is \emph{explodable} if $\eta(G \divideontimes e) \leq \eta(G)-1$.
\begin{theorem}[Meshulam \cite{meshulam2003domination}]\label{th:meshulam}
    For any graph $G$, every edge is either deletable or explodable.
\end{theorem}
This theorem motivates the following definition.
\begin{definition}
    A \emph{deletion-explosion (DE)-sequence} $\pi$ is a sequence of explosions of explodable edges and deletions of deletable edges.
    We denote by $\divideontimes(\pi)$ the set of all edges that are exploded in $\pi$.
    We also write $\pi_{< t}$ to denote the prefix consisting of the first $t-1$ deletions and explosions of $\pi$.
\end{definition}

DE-sequences are a convenient way to obtain lower bounds on $\eta$. 
Notice that $\eta$ is monotonically decreasing during the DE-sequence and strictly decreasing at every explosion.
Therefore, intuitively, if a DE-sequence contains many explosions, then the initial graph has large $\eta$.
However, if a DE-sequence contains many deletions, this might also be helpful for lower-bounding $\eta$ thanks to the following fact.
\begin{proposition}\label{prop:eta-infty}
    If $G$ is a graph containing an isolated vertex, then $\eta(G)= \infty$.
\end{proposition}
A DE-sequence that results in a graph with an isolated vertex is called a \emph{KO-sequence}.
The following is a direct consequence of the above definitions and theorems.
\begin{proposition}\label{prop:explosions-eta}
    Assume that there exists a DE-sequence $\pi$ starting from a graph $H|_Q$ and resulting in a graph $G$.
    Then, $\eta(H|_{Q}) \geq \explosions{\pi} + \eta(G)$.
    Furthermore, if $\pi$ is a KO-sequence, then $\eta(H|_Q) = \infty$.
\end{proposition}

Following \cite{haxell2025improved}, we partition the item set $I$ into \emph{thin} items $J$ and \emph{fat} items $F$. An item $i \in I$ is fat if $a(i) \geq \alpha T$, and thin otherwise.
For any $Q \subseteq P$, let $F_Q:=F \cap \bigcup_{p \in Q} R_p $ denote the fat items in which some players in $Q$ are interested. 
\subsection{Some observations}
We present some preliminary observations that will be useful in the analysis. The first observation due to LP duality.
\begin{observation}\label{obs:lp-duality}
Let $y \in \RR^P$ and $z \in \RR^J$ be a feasible solution to \eqref{lp:dclp} with configuration threshold $T$.
Then, $y(P) \leq z(J)$.
\end{observation}
\begin{proof}
    By definition, \eqref{lp:clp} is feasible for configuration threshold $T$. By LP Duality, \eqref{lp:dclp} must be bounded for $T$.
    However, if $\sum_{p \in P} y_p > \sum_{j \in J} z_j$, then the optimal objective of \eqref{lp:dclp} would be strictly positive.
    In that case, we could scale all variables by an arbitrary factor while maintaining feasibility of the LP. As a result, \eqref{lp:dclp} would be unbounded, a contradiction.
\end{proof}
The second observation concerns fat items.
\begin{restatable}{observation}{obsFatClique}\label{obs:fat-clique}
    Let $f \in F$, and let $v^p \in V(H)$ with $f \in v$.
    Then, for any $w^q$ with $w \neq \{f\}$, we have $(v^p,w^q) \notin E(H)$.
\end{restatable}
\begin{proof}
    Let $v^p \in V(H)$ with $f \in v$.
    Since $f$ is fat, we have $a(f) \geq \alpha T$.
    By the minimality of $v$, this implies $v=\{f\}$.
    Now let $(v^p,w^q) \in E(H)$. We must have $v \cap w \neq \emptyset$.
    Hence, $f \in w$, and by the same argument as before, $w= \{f\}$.
\end{proof}
The remaining observations are bounds on the value of a set of items $v \cup w$ of an edge $(v^p,w^q) \in E(H)$.
\begin{restatable}{observation}{obsOneItemLess}\label{obs:one-item-less}
    For any edge $(v^p,w^q) \in E(H)$ and $j \in v \cup w$, we have $a((v \cup w)-j) < 2 \alpha T$.
\end{restatable}
\begin{proof}
    Without loss of generality, let $j \in v$. Then, $v-j \subset v$ and $w-v \subset w$, thus $a((v \cup w)-j) = a(v-j) + a(w-v) < 2 \alpha T$ by the minimality of $v$ and $w$.
\end{proof}

\begin{restatable}{observation}{obsBoundSmallItem}\label{obs:bound-small-item}
    Let $(v^p,w^q) \in E(H)$ and $i \in v \cup w$. Then, $a(v \cup w) < 2\alpha T + a(i)$.
\end{restatable}
\begin{proof}
    Without loss of generality, let $i \in v$. 
    Since $(v^p,w^q) \in E$, it holds that $v \cap w \neq \emptyset$.
    We have 
    $a(v \cup w) = a(v-i)+a(i)+a(w-v) < 2 \alpha T + a(i)$ by minimality of $v$ and $w$.
\end{proof}

\begin{restatable}{observation}{obsBoundBigItem}\label{obs:bound-big-item}
    Let $(v^p,w^q) \in E(H)$, $i \in v$ and $|v|=k \geq 3$. Then, $a(v \cup w) < 2 \alpha T + \frac{\alpha T - a(i)}{k-2}$.
\end{restatable}

\begin{proof}
Let $j \in v-i$.
    By the minimality of $v$, we have $a(v-\{i,j\}) < \alpha T-a(i)$.
    Therefore, there exists an item $\ell \in v-\{i,j\}$ with $a_\ell < \frac{\alpha T - a(i)}{|v-\{i,j\}|}=\frac{\alpha T -a(i)}{k-2}$.
    We can conclude by using \Cref{obs:bound-small-item}.
\end{proof}
\begin{restatable}{observation}{lemLargeEdge}\label{lem:large-edge}
Let $(v^p,w^q) \in E(H)$ and $|v| = k$ for some $k \geq 2$. Then, $a(v \cup w) \leq (2 + \frac{1}{k-1}) \alpha T$.
\end{restatable}
\begin{proof}
Let $j \in v$. 
By the minimality of $v$, we have $a(v-j) < \alpha T$.
Since $|v-j| = k-1$, this implies the existence of $j' \in v-j$ with $a(j') < \frac{1}{k-1} \alpha T$.
The result follows by \Cref{obs:bound-small-item}.
\end{proof}

\section{An integrality gap of 3.5}
\label{sec:simpler}
For an arbitrary fixed subset $Q \subseteq P$ of players, we will show $\eta(H|_Q) \geq |Q|$ by constructing a DE-sequence $\rho$.
We first introduce some necessary definitions.

Immediately before the $t$-th explosion or deletion in $\rho$, we say that we are at \emph{time} $t$.
Let $H^{(t)} = (V^{(t)},E^{(t)})$ be the remaining graph at time $t$.
Initially, let $H|_Q = H^{(1)}$. 
Let $\pi$ be a DE-sequence starting at a graph $G$ and resulting in a graph $G'$.
    A vertex $v^p$ \emph{survives} $\pi$ if $v^p \in V(G')$.
    A vertex $v^p$ is \emph{exploded} by $\pi$ if $v^p \in V(G)$ but $v^p$ does not survive $\pi$.
As in \cite{haxell2025improved}, we use the notion of ``covers'' to keep track of the impact of a DE-sequence on the graph. We recall that $J$ denotes the thin items. 
\begin{definition}[Cover]
    A \emph{cover} of a DE-sequence $\pi$ is a set $W\subseteq J$ such that for all vertices $v^p$ with $v \subseteq J$ that are exploded by $\pi$, $v \cap W \neq \emptyset$.
    \label{def:cover}
\end{definition}

For example, if $\pi$ consists of exploding an edge $(v^p,w^q)$, where $v \cup w \subseteq J$, then $v \cup w$ is a cover of $\pi$, though this is not necessarily the cover with the minimum value. We also emphasize that a cover consists entirely of thin items. 
Along with our DE-sequence $\rho$, we will maintain a cover $W^{(t)}$ of $\rho_{<t}$ which will be useful in the analysis.

We now describe the construction of our DE-sequence $\rho$. At time $t$, as long as some rule from the following list is applicable, perform the first applicable one in the list.
\begin{itemize}
    \item \textbf{{Rule 1 (fat)}}\label{st:n1}: If there is an edge $e=(v^p,w^q)$ with $(v \cup w) \cap F \neq \emptyset$, perform the DE-sequence $\pi$ that consists of deleting $e$ if $e$ is deletable, and exploding it otherwise. We set $W_\pi = \emptyset$.  
    \item\textbf{Rule 2 (delete)}: If there is a deletable edge $e$, perform the DE-sequence $\pi$ that consists of deleting $e$, and set $W_\pi = \emptyset$.
    \item \textbf{Rule 3 (cheap)}: If there is a DE-sequence $\pi$ starting at $H^{(t)}$ and a set $W_\pi$ with $a(W_\pi) \leq 2 \alpha T \cdot \explosions{\pi}$ such that $W_\pi \cup W^{(t)}$ is a cover of $\pi$, perform the deletions and explosions of $\pi$. 
    \item \textbf{Rule 4 (medium)}: If there is a DE-sequence $\pi$ starting at $H^{(t)}$ with $\explosions{\pi}=3$ which has a cover $W_\pi$ with $|W_\pi| \leq 7$, perform the deletions and explosions of $\pi$.
    \item \textbf{Rule 5 (costly)}: If there is a DE-sequence $\pi$ starting at $H^{(t)}$ with $\explosions{\pi}=2$ which has a cover $W_\pi$ with $|W_\pi| \leq 5$, perform the deletions and explosions of $\pi$.
\item \textbf{Rule 6 (many)}: If there is an edge $e=(v^p,w^q) \in E^{(t)}$ with $|v \cup w| \geq 4$, perform the DE-sequence $\pi$ that consists of simply exploding $e$. Set $W_\pi = v \cup w$.
\end{itemize}
Every time after a rule is applied, we set $W^{(t+|\pi|)}=W^{(t)} \cup W_\pi$.

We note that our procedure to obtain DE-sequences is similar to that used by  \cite{haxell2025improved} \emph{to the extent of} \ruleone to \rulefive. A critical difference is that in their procedure, when 
these five rules 
are no longer applicable, they continue by exploding an arbitrary remaining edge, while our \rulesix imposes a cardinality condition on the edges to be exploded.
Nonetheless, our identical \ruleone to \rulefive allow us to use two of their theorems (see \Cref{lem:no-small-X} and \Cref{lem:lp-application-1}) later on. 

If there is any KO-sequence starting from $H^*$, then \Cref{prop:explosions-eta} immediately implies that $\eta(H|_Q) \geq |Q|$ and we are done. So from now on, we always assume that no KO-sequence starts from $H^*$.

It is easy to see that $\rho$ can be divided into two parts: the prefix is entirely defined by the application of \ruleone and the suffix is entirely defined by the application of \ruletwo to \rulesix. 
Let $\sigma$ be this suffix of $\rho$ and the prefix will be written as $\rho -\sigma$. 

Recall that $F_Q$ is the set of fat items in which some players in $Q$ are interested.
It follows from Observation~\ref{obs:fat-clique} that each fat item in $F_Q$ forms 
a clique in the initial graph $H{|_Q} = H^{(1)}$. Furthermore, all other vertices of $H^{(1)}$ consist entirely of thin items. This implies the following.

\begin{observation}\label{obs:fat-thin}
Suppose that $\rho$ is not a KO-sequence. 
Then 
\begin{enumerate}
    \item[(i)] $\explosions{\rho} \geq \explosions{\sigma} + |F_Q|$. In particular, 
    $\explosions{\rho - \sigma} = |F_Q|$. 
    \item[(ii)] No vertex containing a fat item $H^{(1)}$ survives  $\rho -\sigma$.
    \item[(iii)] If in $H^{(1)}$ a vertex contains a thin item,  this vertex consists entirely of thin items and it survives $\rho -\sigma$. 
\end{enumerate}
\end{observation}

We define $t^*:=|\rho|$ and $H^*=(V^*,E^*):=H^{(t^*+1)}$ as the final graph of the DE-sequence. We also write $W^*:=W^{(t^*+1)}$.
Note that $W^*$ consists entirely of thin items that are added to $W^*$ during the suffix sequence $\sigma$. 
To facilitate notation, we partition our cover $W^*$ into 
\begin{equation}\label{eq:W-decomp}
    W^* = W_{\sf cheap} \sqcup W_{\sf medium} \sqcup W_{\sf costly} \sqcup W_{\sf many}
\end{equation}
according to the rule through which an item was added to $W^*$.
Analogously to the partition of $W$, we denote the number of explosions in $\sigma$ by categories: 
\begin{equation}
\explosions{\sigma} = n_{\sf cheap} + n_{\sf medium}+n_{\sf costly}+n_{\sf many} \label{eq:n-decomp}
\end{equation}
Since $|F_Q|$ is simply the number of explosions in the prefix $\rho-\sigma$ (\Cref{obs:fat-thin}(i)), our final goal is to show that $|Q|-|F_Q|$ is upper-bounded by the number of explosions involving thin items, namely $\explosions{\sigma}$.  
The general idea of the analysis is to prove an upper bound on $a(W^*)$ in terms of $\explosions{\sigma}$ and a lower bound on $a(W^*)$ in terms of $|Q|-|F_Q|$. We start with the upper bound.
\begin{restatable}{lemma}{lemBoundX}\label{lem:X-upper-bound}
    The following bounds hold.
    \begin{enumerate}
        \item[(i)] $a(W_{\sf cheap}) \leq 2 \alpha T \cdot n_{\sf cheap}$
        \item[(ii)] $a(W_{\sf medium}) \leq \frac{7}{3} \alpha T \cdot n_{\sf medium}$
        \item[(iii)] $a(W_{\sf costly}) \leq \frac{5}{2} \alpha T \cdot n_{\sf costly}$
        \item[(iv)] $a(W_{\sf many}) \leq \frac{5}{2} \alpha T \cdot n_{\sf many}$
        \item[(v)] $a(W^*) \leq \frac{5}{2} \alpha T \cdot \explosions{\sigma}$
    \end{enumerate}
\end{restatable}
\begin{proof}

For (i), consider a DE-sequence $\pi$ with $\explosions{\pi}=k$ performed by \rulethree. 
    Then, $a(W_\pi) \leq 2 \alpha T \cdot k$.
    Summing over all applications of \rulethree, we get $a(W_{\sf cheap})
    \leq 2 \alpha T \cdot n_{\sf cheap}$.
    For (ii), observe that for every sequence $\pi$ performed by \rulefour, $\explosions{\pi}=3$ and $|W_\pi| \leq 7$. 
    Therefore, $|W_{\sf medium}| \leq \frac{7}{3} \cdot n_{\sf medium}$.
    Since $\sigma$ cannot contain any explosions that involve fat items by \Cref{obs:fat-thin}, we get $a(W_{\sf medium}) < |W_{\sf medium}| \cdot \alpha T \leq \frac{7}{3} \alpha T \cdot n_{\sf medium}$.
    (iii) has the same proof as (ii). For (iv), consider an explosion of an edge $e=(v^p,w^q)$ by \rulesix, we have either $|v| \geq 3$ or $|w| \geq 3$. By \Cref{lem:large-edge}, we get $a(W_\pi) = a(v \cup w) \leq \frac{5}{2} \alpha T$.
    Summing over all applications of \rulesix gives $a(W_{\sf many}) \leq \frac{5}{2} \alpha T \cdot n_{\sf many}$. 
    (v) follows from (i) to (iv) using \Cref{eq:W-decomp} and \Cref{eq:n-decomp}.
    \end{proof}

\subsection{The structure of the final graph}
The following lemmata describe some properties of the final graph $H^*$, where no more rule can be applied. Our first lemma regards the number of items in a vertex of $H^*$.
\begin{restatable}{lemma}{lemCardTwo}\label{lem:card-two}
For all $v^p \in V^*$, we have $|v| = 2$.
\end{restatable}
\begin{proof}
As all fat items are eliminated by \ruleone according to \Cref{obs:fat-thin}, we have $|w| \geq 2$ for all vertices $w^q \in V^*$. In particular, $|v| \geq 2$. 
Suppose $|v| \geq 3$. Since $\sigma$ is not a KO-sequence, there must be some incident edge $e=(v^p,w^q) \in E^*$. 
If $|v \cap w| \geq 2$ with $i,j \in v \cap w$, let $W=v \cup w$. Then  
$a(W) \leq a(v-i)+a(w-j) < 2 \alpha T$ by the minimality of $v$ and $w$. Furthermore, all vertices in the closed neighborhood 
of $v^p$ and $w^q$ intersect $W$. 
But then, \rulethree is applicable, a contradiction.
If $|v \cap w| = 1$, we have $|v \cup w| = |v|+|w|-1 \geq 3 + 2 - 1 = 4$ and thus \rulesix is applicable, another contradiction.
The proof follows. 
\end{proof}
The next two statements say that any configuration must have a ``large'' intersection with our cover $W^*$. 
\begin{restatable}{observation}{obsVertexExists}\label{obs:vertex-exists}
Let $p \in Q$ and $D \subseteq R_p \cap J$.
\begin{itemize}
    \item  For all $t$, If $D \cap W^{(t)} = \emptyset$ and $a(D) \geq \alpha T$, then there exists $v \subseteq D$ such that $v^p \in V^{(t)}$.
    \item In particular, if $D$ does not contain any vertex from $H^*$, then $a(D - W^*) < \alpha T$.
\end{itemize}
\end{restatable}
\begin{proof}
    We start by removing items from $D$ until the resulting set $v$ is a minimal set of player $p$. Then by definition 
    $v^{p}$ is a minimal vertex of $p$ in $H_{|_Q} = H^{(1)}$ (see Definition~\ref{def:allocationGraph}). 
    Thus, $v^p \in V^{(1)}$.
    By \Cref{obs:fat-thin} and the fact that  $v \subseteq J$, vertex $v^p$ cannot have been exploded by \ruleone.
    But since $v \cap W^{(t)} = \emptyset$, $v^p$ can also not have been exploded in $\sigma_{< t}$.
    Therefore, $v^p$ survives $\rho_{<t}$ and thus $v^p \in V^{(t)}$.
    The second point follows from the contrapositive of the first point with $t = t^* + 1$ and $D=C-W^*$.
\end{proof}
\begin{theorem}[\cite{haxell2025improved}, Theorem 4.1]\label{lem:no-small-X}
    Let $C \in \mathcal{C}(T,p)$ and suppose that \ruleone, \ruletwo and \rulethree are not applicable at time $t$.
    \begin{itemize}
        \item  If $a(C \cap W^{(t)}) < (1-2 \alpha )T$, then \rulefive is applicable at time $t$.
        \item If $a(C \cap W^{(t)}) < (1-3 \alpha )T$, then \rulefour\ is applicable at time $t$.
    \end{itemize}
    
\end{theorem}
In particular, the first point implies that $a(C \cap W^*) \geq (1-2\alpha)T$ for every configuration $C$.

We finish this subsection by observing that we can use \Cref{lem:card-two} to show that $H^*$ is a subgraph of a line graph.
We will see in the next subsection how we can exploit this fact.

Let $K$ be the set of items $i$ such that there exists a vertex $v^p \in V^*$ with $i \in v$.
We claim that $K \cap W^* = \emptyset$, otherwise, \rulethree would be applicable. To see this, consider an edge $e=(v^p,w^q) \in E^*$ with item $i \in v \cap W^*$. 
Let $W=(v\cup w)-i$. Then $a(W) < 2\alpha T$ by \Cref{obs:one-item-less}, implying the applicability of \rulethree. 

We define a graph $B=(K,E')$, 
which has vertex set $K$ and there is an edge $e \in E'$ if $e$ appears as a vertex in $H^*$ (i.e., for some player $p \in Q$ we have $e^p \in V^*$. Recall that in $H^*$, all vertices consist of exactly 2 items by \Cref{lem:card-two}.)
Let $L$ be the line graph of $B$.
\Cref{fig:sets} illustrates the different subsets of $I$.
\begin{figure}[!htbp]
  \centering
  \resizebox{0.7\linewidth}{!}{%
\begin{circuitikz}
\tikzstyle{every node}=[font=\fontsize{14.4pt}{18.7pt}\selectfont]
\draw [ color={rgb,255:red,0; green,0; blue,0}, draw opacity=0.2, fill={rgb,255:red,164; green,207; blue,153}, fill opacity=0.34] (9.375,-4.875) rectangle (21.375,-12.875);
\node[anchor=north, align=center, inner xsep=0.080cm, inner ysep=0.085cm, font=\fontsize{17.28pt}{22.5pt}\selectfont] at (15.375,-5.2) {$J$ (Thin items)};
\draw  (9.375,-4.875) rectangle (25.625,-12.875);
\draw  (21.375,-4.875) rectangle (25.625,-12.875);
\node[anchor=north, align=center, inner xsep=0.080cm, inner ysep=0.085cm, font=\fontsize{17.28pt}{22.5pt}\selectfont] at (23.5,-5.2) {$F$ (fat items)};
\draw  (22,-6.875) rectangle (25,-12);
\node[anchor=north, align=center, inner xsep=0.080cm, inner ysep=0.085cm] at (23.5,-9) {$F_Q$};
\begin{scope}[overlay]
\clip (9.375,-4.875) -- (25.625,-4.875) -- (25.625,-12.875) -- (9.375,-12.875) -- cycle;
\clip (21.375,-4.875) -- (25.625,-4.875) -- (25.625,-12.875) -- (21.375,-12.875) -- cycle;
\fill[fill={rgb,255:red,146; green,189; blue,251}, fill opacity=0.19] (21.1,-13.1) rectangle (25.9,-4.6);
\draw (9.375,-4.875) -- (25.625,-4.875) -- (25.625,-12.875) -- (9.375,-12.875) -- cycle;
\draw (21.375,-4.875) -- (25.625,-4.875) -- (25.625,-12.875) -- (21.375,-12.875) -- cycle;
\end{scope}
\begin{scope}[overlay]
\clip (9.375,-4.875) -- (25.625,-4.875) -- (25.625,-12.875) -- (9.375,-12.875) -- cycle;
\begin{scope}[even odd rule]
\clip (9.1,-13.1) rectangle (25.9,-4.6) (21.375,-4.875) -- (25.625,-4.875) -- (25.625,-12.875) -- (21.375,-12.875) -- cycle;
\fill[fill={rgb,255:red,38; green,235; blue,57}, fill opacity=0.19] (9.1,-13.1) rectangle (25.9,-4.6);
\draw (9.375,-4.875) -- (25.625,-4.875) -- (25.625,-12.875) -- (9.375,-12.875) -- cycle;
\draw (21.375,-4.875) -- (25.625,-4.875) -- (25.625,-12.875) -- (21.375,-12.875) -- cycle;
\end{scope}
\end{scope}
\begin{scope}[overlay]
\clip (9.375,-4.875) -- (25.625,-4.875) -- (25.625,-12.875) -- (9.375,-12.875) -- cycle;
\clip (9.375,-6.125) -- (17.75,-6.125) -- (17.75,-12.875) -- (9.375,-12.875) -- cycle;
\fill[fill={rgb,255:red,235; green,199; blue,38}, fill opacity=0.19] (9.1,-13.1) rectangle (18,-5.9);
\draw (9.375,-4.875) -- (25.625,-4.875) -- (25.625,-12.875) -- (9.375,-12.875) -- cycle;
\draw (9.375,-6.125) -- (17.75,-6.125) -- (17.75,-12.875) -- (9.375,-12.875) -- cycle;
\end{scope}
\begin{scope}[overlay]
\clip (9.375,-4.875) -- (25.625,-4.875) -- (25.625,-12.875) -- (9.375,-12.875) -- cycle;
\clip (17.75,-6.125) -- (21.375,-6.125) -- (21.375,-12.875) -- (17.75,-12.875) -- cycle;
\fill[fill={rgb,255:red,217; green,17; blue,17}, fill opacity=0.35] (17.5,-13.1) rectangle (21.6,-5.9);
\draw (9.375,-4.875) -- (25.625,-4.875) -- (25.625,-12.875) -- (9.375,-12.875) -- cycle;
\draw (17.75,-6.125) -- (21.375,-6.125) -- (21.375,-12.875) -- (17.75,-12.875) -- cycle;
\end{scope}
\draw [ color={rgb,255:red,0; green,0; blue,0}, draw opacity=0 ] (9.375,-6.125) rectangle (11.375,-12.875);
\node[anchor=center, align=center] at (10.375,-9.5) {$W_{\sf cheap}$};
\begin{scope}[overlay]
\clip (9.375,-4.875) -- (25.625,-4.875) -- (25.625,-12.875) -- (9.375,-12.875) -- cycle;
\clip (9.375,-6.125) -- (17.75,-6.125) -- (17.75,-12.875) -- (9.375,-12.875) -- cycle;
\clip (9.375,-6.125) -- (11.375,-6.125) -- (11.375,-12.875) -- (9.375,-12.875) -- cycle;
\fill[fill={rgb,255:red,28; green,112; blue,28}, fill opacity=0] (9.1,-13.1) rectangle (11.6,-5.9);
\draw (9.375,-4.875) -- (25.625,-4.875) -- (25.625,-12.875) -- (9.375,-12.875) -- cycle;
\draw (9.375,-6.125) -- (17.75,-6.125) -- (17.75,-12.875) -- (9.375,-12.875) -- cycle;
\draw[color={rgb,255:red,0; green,0; blue,0}, draw opacity=0] (9.375,-6.125) -- (11.375,-6.125) -- (11.375,-12.875) -- (9.375,-12.875) -- cycle;
\end{scope}
\draw  (17.75,-6.125) rectangle (21.375,-12.875);
\node[anchor=center, align=center] at (19.5625,-9.5) {$K$ (remaining\\ items in $H^*$)};
\begin{scope}[overlay]
\clip (9.375,-4.875) -- (25.625,-4.875) -- (25.625,-12.875) -- (9.375,-12.875) -- cycle;
\clip (9.375,-6.125) -- (17.75,-6.125) -- (17.75,-12.875) -- (9.375,-12.875) -- cycle;
\clip (13.5,-6.125) -- (15.625,-6.125) -- (15.625,-12.875) -- (13.5,-12.875) -- cycle;
\fill[fill={rgb,255:red,225; green,168; blue,82}, fill opacity=0] (13.3,-13.1) rectangle (15.9,-5.9);
\draw (9.375,-4.875) -- (25.625,-4.875) -- (25.625,-12.875) -- (9.375,-12.875) -- cycle;
\draw (9.375,-6.125) -- (17.75,-6.125) -- (17.75,-12.875) -- (9.375,-12.875) -- cycle;
\draw[color={rgb,255:red,0; green,0; blue,0}, draw opacity=0] (13.5,-6.125) -- (15.625,-6.125) -- (15.625,-12.875) -- (13.5,-12.875) -- cycle;
\end{scope}
\begin{scope}[overlay]
\clip (9.375,-4.875) -- (25.625,-4.875) -- (25.625,-12.875) -- (9.375,-12.875) -- cycle;
\begin{scope}[even odd rule]
\clip (9.1,-13.1) rectangle (25.9,-4.6) (21.375,-4.875) -- (25.625,-4.875) -- (25.625,-12.875) -- (21.375,-12.875) -- cycle
(9.375,-6.125) -- (17.75,-6.125) -- (17.75,-12.875) -- (9.375,-12.875) -- cycle;
\clip (9.1,-13.1) rectangle (25.9,-4.6) (17.75,-6.125) -- (21.375,-6.125) -- (21.375,-12.875) -- (17.75,-12.875) -- cycle
(9.375,-6.125) -- (11.375,-6.125) -- (11.375,-12.875) -- (9.375,-12.875) -- cycle;
\clip (9.1,-13.1) rectangle (25.9,-4.6) (11.375,-6.125) -- (13.5,-6.125) -- (13.5,-12.875) -- (11.375,-12.875) -- cycle;
\clip (9.1,-13.1) rectangle (25.9,-4.6) (13.5,-6.125) -- (15.625,-6.125) -- (15.625,-12.875) -- (13.5,-12.875) -- cycle;
\clip (9.1,-13.1) rectangle (25.9,-4.6) (15.625,-6.125) -- (17.75,-6.125) -- (17.75,-12.875) -- (15.625,-12.875) -- cycle;
\fill[fill={rgb,255:red,184; green,154; blue,237}, fill opacity=0.09] (9.1,-13.1) rectangle (25.9,-4.6);
\draw (9.375,-4.875) -- (25.625,-4.875) -- (25.625,-12.875) -- (9.375,-12.875) -- cycle;
\draw (21.375,-4.875) -- (25.625,-4.875) -- (25.625,-12.875) -- (21.375,-12.875) -- cycle;
\draw (9.375,-6.125) -- (17.75,-6.125) -- (17.75,-12.875) -- (9.375,-12.875) -- cycle;
\draw (17.75,-6.125) -- (21.375,-6.125) -- (21.375,-12.875) -- (17.75,-12.875) -- cycle;
\draw[color={rgb,255:red,0; green,0; blue,0}, draw opacity=0.2] (9.375,-6.125) -- (11.375,-6.125) -- (11.375,-12.875) -- (9.375,-12.875) -- cycle;
\draw[color={rgb,255:red,0; green,0; blue,0}, draw opacity=0.2] (11.375,-6.125) -- (13.5,-6.125) -- (13.5,-12.875) -- (11.375,-12.875) -- cycle;
\draw[color={rgb,255:red,0; green,0; blue,0}, draw opacity=0.2] (13.5,-6.125) -- (15.625,-6.125) -- (15.625,-12.875) -- (13.5,-12.875) -- cycle;
\draw[color={rgb,255:red,0; green,0; blue,0}, draw opacity=0] (15.625,-6.125) -- (17.75,-6.125) -- (17.75,-12.875) -- (15.625,-12.875) -- cycle;
\end{scope}
\end{scope}
\draw [ color={rgb,255:red,0; green,0; blue,0}, draw opacity=0 ] (15.625,-6.125) rectangle (17.75,-12.875);
\node[anchor=center, align=center] at (16.6875,-9.5) {$W_{\sf many}$};
\begin{scope}[overlay]
\clip (9.375,-4.875) -- (25.625,-4.875) -- (25.625,-12.875) -- (9.375,-12.875) -- cycle;
\clip (9.375,-6.125) -- (17.75,-6.125) -- (17.75,-12.875) -- (9.375,-12.875) -- cycle;
\clip (15.625,-6.125) -- (17.75,-6.125) -- (17.75,-12.875) -- (15.625,-12.875) -- cycle;
\fill[fill={rgb,255:red,224; green,81; blue,81}, fill opacity=0.] (15.4,-13.1) rectangle (18,-5.9);
\draw (9.375,-4.875) -- (25.625,-4.875) -- (25.625,-12.875) -- (9.375,-12.875) -- cycle;
\draw (9.375,-6.125) -- (17.75,-6.125) -- (17.75,-12.875) -- (9.375,-12.875) -- cycle;
\draw[color={rgb,255:red,0; green,0; blue,0}, draw opacity=0] (15.625,-6.125) -- (17.75,-6.125) -- (17.75,-12.875) -- (15.625,-12.875) -- cycle;
\end{scope}
\node [font=\fontsize{17.28pt}{22.5pt}\selectfont] at (17.5,-4.125) {$I$ (all items)};
\draw [ color={rgb,255:red,0; green,0; blue,0}, draw opacity=0.2 ] (13.5,-6.125) rectangle (15.625,-12.875);
\node[anchor=center, align=center] at (14.5625,-9.5) {$W_{\sf costly}$};
\draw [ color={rgb,255:red,0; green,0; blue,0}, draw opacity=0.2 ] (11.375,-6.125) rectangle (13.5,-12.875);
\node[anchor=center, align=center] at (12.4375,-9.5) {$W_{\sf medium}$};
\draw  (9.375,-6.125) rectangle (17.75,-12.875);
\node[anchor=south, align=center, inner xsep=0.080cm, inner ysep=0.085cm] at (13.5625,-12.725) {$W^*$ (cover)};
\end{circuitikz}
  }%
  \caption{An illustration of the different sets of items that we define.
Explosions of \ruleone \ only involve items in $F_Q \subseteq F$. 
The vertices of all other explosions are covered by $W^*$.
Vertices in $V^*$ only consist of items in $K$.
Finally, there might exist $i \in I-F_Q-W^*-K$ that appears neither in $\rho$ nor in $V^*$, if no vertex including $i$ is minimal, or if $i \notin \bigcup_{q \in Q} R_q$.
}
    \label{fig:sets}
\end{figure}
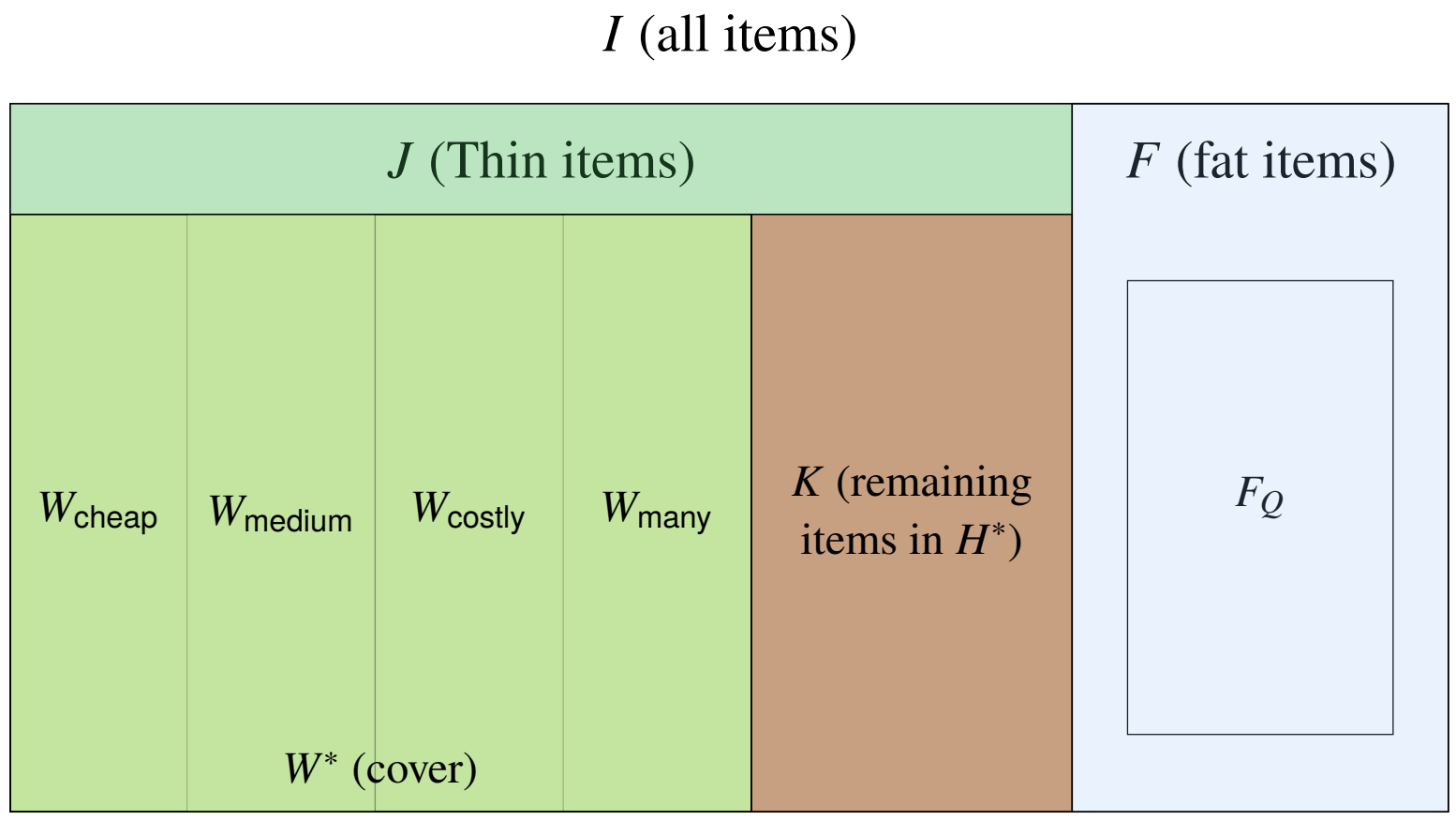
\begin{restatable}{lemma}{lemLineGraph}\label{obs:line-graph}
    There is a (potentially empty) set of edges in $L$ whose removal makes $L$ isomorphic to $H^*$.\footnote{The reason why we cannot claim that $L$ is isomorphic to $H^*$ is that some of the edges of $L$ might have been deleted by  the DE-sequence $\rho$ when \ruletwo is applied.}
\end{restatable}
\begin{proof}
    Let $\overline{V^*} = \{v \mid v^p \in V^*\}$.
    By construction of $E'$, we have $E' \subseteq \overline{V^*}$. 
    For every $v \in \overline{V^*}$, we have $v \subseteq K$ and $|v|=2$ by \Cref{lem:card-two}, thus $v \in E'$.
    We obtain $\overline{V^*}=E'$.

    Furthermore, if $e=(v^p,w^q) \in E^*$, then $v \cap w \neq \emptyset$.
    Let $j \in v \cap w$.
    In $B$, both $v$ and $w$ are edges with common endpoint $j$.
    Therefore, $(v,w) \in E(L)$. The result follows. 
\end{proof}

\subsection{Another lower bound on $\eta$}
We would like to exploit the structure of $H^*$ in order to obtain a better lower bound on $\eta(H^*)$.
To this end, we use a theorem by Aharoni, Berger and Meshulam. The theorem gives a lower bound on $\eta$ in terms of a graph parameter $\Gamma$. 
Let $G=(V,E)$ be a graph.
A \emph{vector representation} $S: V \rightarrow \RR^\ell$ of $G$ assigns a vector of fixed dimension $\ell$ to every vertex of $G$, such that $S(u)\cdot S(v) \geq 0$ for all $u,v$ and furthermore, $S(u) \cdot S(v)  \geq 1$ if $(u,v) \in E$.
For example, choosing $S(v)$ as the edge indicence vector of $v$ makes $S$ a vector representation of $G$.
A nonnegative vector $\beta \in \RR^V$ is \emph{dominating} $S$ if $\sum_{v \in V} \beta_v(S(v)\cdot S(u)) \geq 1$ for all $u \in V$.
Here, $\cdot$ denotes the dot product while $\beta_v$ is a simple scalar.
The parameter $\Gamma$ is defined as 
$$\Gamma(G)=\sup_{S}\min_\beta \beta(V),$$
where the supremum is over all vector representations $S$ of $G$ and the minimum is over all nonnegative vectors dominating $S$.
\begin{theorem}[Aharoni, Berger and Meshulam~\cite{aharoni2005eigenvalues}]\label{th:aharoni}
    For any graph $G$, it holds that $\eta(G) \geq \Gamma(G)$.
\end{theorem}
\begin{definition}[Fractional edge dominating set] 
A \emph{fractional edge dominating set}\footnote{Originally, the authors of \cite{aharoni2005eigenvalues} called this parameter \emph{fractional cover}. However, we choose to use the term \emph{fractional edge dominating set} to avoid confusion with the cover of a DE-sequence and the fractional vertex cover of a graph.} of a graph $G=(V,E)$ is a vector $r \in (\RR^+)^E$ with the property that for all $e \in E$, $\sum_{f \in E} |e \cap f|r_f \geq 1$. 
    Let $\omega^*(G)=\min\{r(E) \mid \text{$r$ is a fractional edge dominating set of $G$}\}$.
\end{definition}
This parameter gives a lower bound on $\Gamma$, as observed in \cite{aharoni2005eigenvalues}. 
We include a proof for completeness.
\begin{restatable}[{Aharoni, Berger, Meshulam~\cite{aharoni2005eigenvalues}}]{proposition}{propOmega}\label{obs:aharoni}
    If $B$ is a graph and $L$ its line graph, $\Gamma(L) \geq \omega^*(B)$.
\end{restatable}
\begin{proof}
    For any $e \in E(B)$, let $S(e) \in \RR^{V(B)}$ be the vertex incidence vector of $e$.
    Then, $S$ is a vector representation for $L$.
    Let $\beta \in \RR^{V(L)}$ be a vector that dominates $S$.
    Setting $r=\beta$, we claim that $r$ is a fractional edge dominating set of $B$. Indeed, for every $e \in E(B)$, $$\sum_{f \in E(B)} |e \cap f|r_f =\sum_{v \in V(L)} \beta_v|e \cap f| = \sum_{v \in V(L)} \beta_v(S(e) \cdot S(v))\geq 1,$$
    as $E(B)=V(L)$ and $S(u) \cdot S(v) = |e \cap f|$.
    Therefore, for our chosen $S$, we have that $\min_\beta \beta(V(L))$ can not be smaller than $\omega^*(B)$.
    It follows that $\Gamma(L) \geq \omega^*(B)$.
\end{proof}

\Cref{obs:aharoni} can be used to prove a consequence of \Cref{th:aharoni} that will be useful in our context.
A similar statement was already observed by Aharoni, Berger, Guo and Kotlar~\cite{aharoni2025coloring}.
\begin{restatable}{proposition}{thFracVc}\label{th:frac-vc}
    Let $B$ be a graph and $L$ be its line graph.
    Let $L'$ be obtained from $L$ by removing a (potentially empty) set of edges.
    We have $\tau^*(B) \leq 2 \cdot \eta(L')$.
\end{restatable}
\begin{proof}
    By \Cref{th:aharoni}, $\eta(L') \geq \Gamma(L')$.
    Let $S$ be any vector representation of $L$. Since $L'$ is obtained from $L$ by removing edges, $S$ is also a valid vector representation for $L'$.
    Furthermore, whether a vector $\beta$ dominates $S$ does not depend on the edge set in the underlying graph.
    Consequently, $\Gamma(L') \geq \Gamma(L) \geq \omega^*(B)$, where the last inequality holds by \Cref{obs:aharoni}.
    
    We now show $\omega^*(B) \geq \frac{\tau^*(B)}{2}$.
    Given a fractional edge dominating set $r \in \RR^{E(B)}$ with $r(E(B))=\omega^*(B)$, we set $z'_v = \sum_{e \in \delta(v)} r_e$ for all $v \in V(B)$.
    We have $$z'(V(B)) = \sum_{v \in V(B)}\sum_{e \in \delta(v)} r_e = \sum_{e \in E(B)} \sum_{v \in e} r_e = \sum_{e \in E(B)} 2r_e = 2 \cdot r(E(B)).$$
    The resulting $z'$ is a fractional vertex cover since for every edge $e=(u,v)$, we have $$z'_u + z'_v = \sum_{f \in \delta(u)} r_f + \sum_{f \in \delta(v)} r_f = \sum_{f \in E(B)} |e \cap f|r_f \geq 1.$$
    In summary, we get $$\tau^*(B) \leq z'(V(B)) = 2r(E(B))=2\omega^*(B) \leq 2\Gamma(L) \leq 2\Gamma(L') \leq 2\eta(L').$$
\end{proof}

\subsection{Lower-bounding the value of the cover}
We now give an assignment to \eqref{lp:dclp} to obtain a lower bound on $a(W^*)$ using \Cref{obs:lp-duality}.
Let $z'$ be a minimum fractional vertex cover of $B$. For all $p \in P$ and $i \in I$, we set
\[
\begin{minipage}{0.45\textwidth}
\begin{equation*}
y_p = \begin{cases}
(1-\alpha) T & \text{if $p \in Q$}\\
0 & \text{otherwise.}
\end{cases}\end{equation*}
\end{minipage}
\qquad
\begin{minipage}{0.45\textwidth}
\begin{equation*}
    z_i = \begin{cases}
(1-\alpha) T & \text{if $i \in F_Q$}\\
a(i) &  \text{if $i \in W^*$}\\
z'_i \cdot \alpha T & \text{if $i \in K$}\\
0 & \text{otherwise}
\end{cases}
\end{equation*}
\end{minipage}
\]
We verify that all dual constraints are satisfied. 

\begin{lemma}\label{lem:dual-feasible}
$(y,z)$ is a feasible solution to \eqref{lp:dclp}.
\end{lemma}
\begin{proof}
    By definition, $y$ and $z$ are non-negative.
Let $p \in P$ and $C \in \mathcal{C}(T,p)$.
By definition of $\mathcal{C}(T,p)$, we know $a(C) \geq T$.
We prove that $z(C) \geq y_p$.
For $p \notin Q$, we have $z(C) \geq 0 = y_p$.
If there exists a fat item $i \in C \cap F_Q$, then $z(C) \geq z_i = (1-\alpha) T = y_p$.
So we can assume $p \in Q$ and $C \subseteq J$.

If there is no vertex $v^p \in V^*$ with $v \subseteq C$, then $a(C-W^*) < \alpha T$ by \Cref{obs:vertex-exists} and thus $$z(C) \geq z(C \cap W^*) = a(C \cap W^*) = a(C)-a(C-W^*) > T - \alpha T = y_p.$$

So we can assume that there exists a vertex $v^p \in V^*$ with $v \subseteq C$.
\Cref{lem:no-small-X} states that if $a(C \cap W^*)<(1-2\alpha)T$, then \rulefive would be applicable and hence $a(C \cap W^*) \geq (1-2\alpha)T$.
Since $v^p \in V^*$ and thus $v \subseteq K$, we must have $z(v) = \alpha T \cdot z'(v) \geq \alpha T$, as $z'$ is a fractional vertex cover of $B$ and $v$ is an edge in $B$ by \Cref{obs:line-graph}.
In total, 
\begin{align*}
    z(C) &= z(C \cap W^*) + z(C - W^*) \\
    &\geq z(C \cap W^*) + z(v) \\
    &\geq (1-2\alpha)T + \alpha T \\
    &= y_p.
\end{align*}
where the first inequality holds because $v \subseteq C-W^*$.

\end{proof}
We can now obtain our lower bound on $a(W^*)$ using LP duality.
\begin{lemma}\label{lem:lp-duality-application}
$|F_Q|(1-\alpha)T + a(W^*) + 2\eta(H^*)\alpha T \geq |Q| \cdot (1- \alpha) T $.
\end{lemma}
\begin{proof}
    On the one hand, since $K \subseteq J-W^*$ and $z(J-W^*-K)=0$, 
    \begin{align*}
        z(I) &= z(F_Q) + z(W^*) + z(J-W^*) \\
        &= |F_Q| \cdot (1-\alpha) T + a(W^*) + z(K) \\
        &=  |F_Q| \cdot (1-\alpha) T + a(W^*) + \alpha T \cdot z'(K) \\
        &\leq |F_Q| \cdot (1-\alpha) T + a(W^*) + 2\eta(H^*)\cdot \alpha T
    \end{align*}
    where the inequality holds by \Cref{th:frac-vc} and \Cref{obs:line-graph}.
    On the other hand $$y(P) = y(Q) = |Q| \cdot (1-\alpha) T.$$ 

    The proof follows since $(y,z)$ is a feasible to \eqref{lp:dclp}
    (\Cref{lem:dual-feasible}) and 
    $z(I) \geq y(P)$
    (\Cref{obs:lp-duality}). 
    \end{proof}
\begin{lemma}\label{lem:combining}
For $\alpha = \frac{2}{7}$, it holds that $\eta(H|_Q) \geq |Q|$.
\end{lemma}
\begin{proof}  
\begin{align*}
    \eta(H|_Q) \cdot (1-\alpha)T
    &\geq (\explosions{\rho} + \eta(H^*)) \cdot (1-\alpha)T\\
    &\geq (|F_Q| + \explosions{\sigma} + \eta(H^*)) \cdot (1-\alpha)T\\
    &\geq a(W^*) + (|F_Q|+ \eta(H^*)) \cdot (1-\alpha)T\\
    &\geq a(W^*) + |F_Q| \cdot (1-\alpha)T + \eta(H^*) \cdot 2 \alpha T\\
    &\geq |Q| \cdot (1-\alpha)T
\end{align*}
where the first inequality uses \Cref{prop:explosions-eta}, the second inequality uses \Cref{obs:fat-thin}, the third inequality holds by \Cref{lem:X-upper-bound} and the fact that $\alpha = \frac{2}{7}$, the fourth inequality again uses 
the fact that $\alpha = \frac{2}{7}$, and the last inequality uses \Cref{lem:lp-duality-application}. The proof follows.
\end{proof}
Since $Q$ is an arbitrary subset of $P$, we can conclude by invoking \Cref{lem:combining} and \Cref{th:hall-general}. 
\begin{theorem}
    The integrality gap of \eqref{lp:clp} is at most $\alpha^{-1}=3.5$.
\end{theorem}

\section{Improving the analysis}
\label{sec:fat}
In this section, we show how to improve the bound of $3.5$. 
In particular, we will show that $H_{\mathcal{I}}(\alpha)$ admits an independent transversal for $\alpha = \frac{5}{17}$.
Our argument is based on the same procedure for constructing a DE-sequence $\rho$ as in the previous section.
To bound $|Q|-|F_Q|$, as we mentioned in the introduction, we will employ two snapshots. The first snapshot is already provided by Haxell and Szab\'{o}.
\begin{lemma}[\cite{haxell2025improved}, Lemma 5.2]
\label{lem:lp-application-1}
    $|Q|-|F_Q| \leq \frac{2\alpha}{1-3\alpha}n_{\sf cheap} + \frac{7}{3}n_{\sf medium}$.
\end{lemma}
To see why this snapshot is helpful, observe that in Lemma~\ref{lem:X-upper-bound}, the worst ratio of the final cover $a(W^*)$ against the total number of explosions in $\sigma$ is attained when all explosions in $\sigma$ are caused by \rulefive and \rulesix. This ratio is one of the bottlenecks in the analysis in \Cref{sec:simpler} (see Lemma~\ref{lem:combining}). The above snapshot implies that Rules~3 and 4 must have been applied a certain number of times. Thus, the ratio $\frac{a(W^*)}{\explosions{\sigma}}$ cannot be as bad as $\frac{5}{2} \alpha T$. 

Our second snapshot is taken when the entire explosion process stops, as in the analysis of the previous section. 
Recall that in the previous section, the analysis relies on upper- and lower-bounding $a(W^*)$. 
To bring these two bounds closer, 
our idea is to construct an artificial version of the value function. Precisely, we will construct an \emph{augmented} value function $a^+$ and show that $a^+(W^*)$ can still be upper-bounded as $a(W^*)$ with a small additional penalty, but its lower bound is improved significantly.
Let
$$a^+(i) = \begin{cases}
    \alpha T & \text{if $i \in W_{\sf medium} \cup W_{\sf costly}$}\\
    \frac{15}{14} \cdot a(i) & \text{if $i \in W_{\sf cheap}$}\\
    \min(\frac{3}{7} \alpha T, \frac{15}{14} \cdot a(i)) & \text{if $i \in W_{\sf many}$ and $a(i) \leq \frac{3}{7} \alpha T$} \\
    a(i) & \text{otherwise.}
\end{cases}$$
\begin{restatable}{lemma}{lemManyBoundPlus}\label{lem:a+many-items-bound}
    For any edge $e=(v^p,w^q) \in E(H)$ exploded by \rulesix, $a^+(v \cup w) \leq \frac{5}{2} \alpha T$.
\end{restatable}
\begin{proof}
First of all, we claim that $v \cup w \subseteq W_{\sf many}$. Suppose not.  Then letting $t$ be the time when $e$ is exploded, there must exist some $i \in v \cup w$ with $i \in W^{(t)}$. But then, $(v \cup w)-i$ is a cover of value $<2 \alpha T$ by \Cref{obs:one-item-less}, and thus $e$ should have been exploded by \rulethree instead of \rulesix, a contradiction.

     Now, without loss of generality, let $|v| \geq |w|$.
     As $e$ was exploded by \rulesix, we must have $|v \cup w| \geq 4$ and $v \cap w \neq \emptyset$.
     Therefore, $|v| \geq 3$.
 If $|v| \geq 4$, then $a(v \cup w) \leq \frac{7}{3} \alpha T$ by \Cref{lem:large-edge}, therefore 
 \begin{equation}\label{eq:7/3}
     a^+(v \cup w) \leq \frac{15}{14} \cdot a(v \cup w) \leq \frac{15}{14} \cdot \frac{7}{3} \alpha T = \frac{5}{2} \alpha T.
 \end{equation}
 From now on, we assume that $|v| = 3$. By our assumption $|v| \geq |w|$ and since $v \cap w \neq \emptyset$, we have $|v \cup w| \leq 5$.
   Let $u \subseteq v \cup w$ be the set of items $i$ in $v \cup w$ with $a(i) \leq \frac{3}{7} \alpha T$.
   Notice that $a^+(u) \leq |u| \cdot \frac{3}{7} \alpha T$ and $a^+((v \cup w)-u)=a((v \cup w)-u)$.
   In particular, notice that this gives us the bound 
   $$a^+(v \cup w)=a(v \cup w)+a^+(v \cup w)-a(v \cup w) \leq a(v \cup w) +a^+(u)-a(u) \leq a(v \cup w) +\frac{1}{14}a(u),$$ where the first inequality uses $a^+((v \cup w)-u)-a((v \cup w)-u) =0$ and the second inequality uses $u \subseteq W_{\sf many}$.
   Furthermore, notice that for all items $i \in v \cup w$, we can assume $a(i) \leq \frac{2}{3} \alpha T$, otherwise $a(v \cup w) \leq \frac{7}{3} \alpha T$ by \Cref{obs:bound-big-item} and we can conclude by \Cref{eq:7/3}.
   We do a case analysis based on the cardinality of $u$.
   \begin{itemize}
    \item  Suppose that $u = \emptyset$. Then $a^+(v \cup w) = a(v \cup w) \leq \frac{5}{2} \alpha T$ by \Cref{lem:large-edge}.
    \item Suppose that $|u| \geq 4$. Then
    \begin{align*}
        a^+(v \cup w) \leq |u| \cdot \frac{3}{7}\alpha T + |(v \cup w)-u| \cdot \frac{2}{3} \alpha T 
        \leq \frac{12}{7} \alpha T + \frac{2}{3} \alpha T
        < \frac{5}{2} \alpha T
    \end{align*}
    where the second inequality uses $|v \cup w| \leq 5$, $|u| \geq 4$ and $\frac{3}{7} \leq \frac{2}{3}$.
   \item Suppose that $1 \leq |u| \leq 2$. Then there exists an item $i \in v \cup w$ with $a(i) \leq \frac{3}{7} \alpha T$. Therefore, $a(v \cup w) < \frac{17}{7} \alpha T$ by \Cref{obs:bound-small-item}. We have
   $$a^+(v \cup w) \leq a(v \cup w)+\frac{1}{14} \cdot a(u)  \leq (\frac{17}{7} +\frac{1}{14} \cdot \frac{6}{7}) \alpha T   < \frac{5}{2} \alpha T$$
   where the second inequality uses $|u| \leq 2$ and $a(i) \leq \frac{3}{7} \alpha T$ for all $i \in u$.
   \item Suppose that $|u|=3$.
   Let $i$ be the item in $v \cup w$ minimizing $a(i)$.

   \begin{itemize}

   \item Suppose that 
$a(i) \geq \frac{2}{5} \alpha T$.
   If $|w| = 2$, then $|(v \cup w)-u|=1$ and thus $a((v \cup w)-u) < \alpha T$ since \ruleone is not applicable.
   If $|w|=3$, then $|(v \cup w)-u|=2$ and no $j \in v \cup w$ can have $a(j) \geq \frac{3}{5} \alpha T$ by minimality of $v$ and $w$. Therefore, $a((v \cup w)-u) \leq 2 \cdot \frac{3}{5} \alpha T$.
   In both cases, we obtain $a((v \cup w)-u) \leq \frac{6}{5} \alpha T$.
   Therefore,
   $$a^+(v \cup w) \leq a^+(u) + a((v \cup w)-u) \leq 3 \cdot \frac{3}{7} \alpha T + \frac{6}{5} \alpha T < \frac{5}{2} \alpha T.$$
   \item Suppose that $a(i)<\frac{2}{5} \alpha T$. By \Cref{obs:bound-small-item}, $a(v \cup w) < \frac{12}{5} \alpha T$.
   Therefore,
   $$a^+(v \cup w) \leq a(v \cup w) + \frac{1}{14} a(u) < \frac{12}{5} \alpha T + \frac{1}{14} \cdot \frac{9}{7} \alpha T  < \frac{5}{2} \alpha T,$$
   where the second inequality uses the facts that $|u| \leq 3$ and that $a(i) \leq \frac{3}{7} \alpha T$ for all $i \in u$.   
   
   \end{itemize}
\end{itemize}

   The proof follows. 
\end{proof}
Using the above lemma, we can prove the equivalent of \Cref{lem:X-upper-bound} for our augmented value function.
\begin{restatable}{lemma}{lemBoundXPlus}\label{lem:a+bound}
    The following bounds hold.
    \begin{enumerate}
        \item[(i)] $a^+(W_{\sf cheap}) \leq \frac{15}{7} \alpha T \cdot n_{\sf cheap}$
        \item[(ii)] $a^+(W_{\sf medium}) \leq \frac{7}{3} \alpha T \cdot n_{\sf medium}$
        \item[(iii)] $a^+(W_{\sf costly}) \leq \frac{5}{2} \alpha T \cdot n_{\sf costly}$
        \item[(iv)] $a^+(W_{\sf many}) \leq \frac{5}{2} \alpha T \cdot n_{\sf many}$
    \end{enumerate}
\end{restatable}
\begin{proof}
For (i), observe that $a^+(W_{\sf cheap}) \leq \frac{15}{14} \cdot a(W_{\sf cheap}) \leq \frac{15}{14} \cdot 2 \alpha T \cdot n_{\sf cheap} = \frac{15}{7}n_{\sf cheap}$ using \Cref{lem:X-upper-bound}.
        For (ii), as already observed in \Cref{lem:X-upper-bound}, we have $\frac{|W_{\sf medium}|}{n_{\sf medium}} \leq \frac{7}{3}$. Therefore, $a^+(W_{\sf medium}) = |W_{\sf medium}| \cdot \alpha T \leq \frac{7}{3} \alpha T$.
        (iii) holds by the same argument. 
        (iv) directly follows from summing over the bound from \Cref{lem:a+many-items-bound} for all edges exploded by \rulesix.
\end{proof}

Now, we define a solution to \eqref{lp:dclp}.
From now on, we will use our particular choice $\alpha = \frac{5}{17}$.
Let $M:=(1- \frac{34}{35} \alpha) T$.
Observe that $M=\frac{5}{7} T$.
For all $p \in P$ and $i \in I$, we set 
\[
\begin{minipage}{0.45\textwidth}
\begin{equation*}
    y_p = \begin{cases}
     M& \text{if $p \in Q$}\\
    0 & \text{otherwise}
\end{cases}
\end{equation*}
\end{minipage}
\qquad
\begin{minipage}{0.45\textwidth}
\begin{equation*}
    z_i=\begin{cases}
    M & \text{if $i \in F$}\\
    a^+(i) & \text{if $i \in W^*$} \\
    z'_i \cdot \alpha T & \text{if $i \in K$}\\
    0 & \text{otherwise.}
\end{cases}
\end{equation*}
\end{minipage}
\]
\begin{restatable}{lemma}{lemDualFeasibleTwo}\label{lem:dual-feasible-2}
    For $\alpha = \frac{5}{17}$, $(y,z)$ is a feasible solution to \eqref{lp:dclp}.
\end{restatable}
\begin{proof}
We define $W_{\sf card}:=W_{\sf medium} \cup W_{\sf costly}$ and $W_{\sf rest} = W_{\sf cheap} \cup W_{\sf many}$. Let $C \in \mathcal{C}(T,p)$.
    As in the proofs of \Cref{lem:dual-feasible} and \Cref{lem:lp-application-1}, we may assume $C \subseteq J$ and $p \in Q$.
We distinguish cases on whether there exists some $v \subseteq C$ with $v^p \in V^*$.

    \begin{itemize} 
    \item Suppose that there is no vertex $v \subseteq C$ with $v^p \in V^*$. Then $a(C \cap W^*)=a(C)-a(C-W^*) \geq (1-\alpha)T$ by \Cref{obs:vertex-exists}.
    Using \Cref{lem:no-small-X}, we also have
    \begin{equation}
            a(C \cap W^*) \geq a(C \cap (W_{\sf cheap} \cup W_{\sf card})) \geq a(C \cap W^{(t')}) \geq (1-2\alpha)T, \label{eq:1-2alphaT}
    \end{equation}
    where $t'$ denotes the first time no rule from \ruleone up to \rulefive is applicable.

    We distinguish cases based on the cardinality of $C \cap W_{\sf card}$.
    \begin{itemize}
        \item Suppose that $|C \cap W_{\sf card}| \leq 1$.
    Then, setting $S:= C \cap W_{\sf cheap}$, we have $$a(S) \geq a(C \cap W^*) - a(C \cap W_{\sf card}) \geq (1-2 \alpha )T - |C \cap W_{\sf card}| \cdot \alpha T \geq (1-2\alpha)T - \alpha T = \frac{2}{5} \alpha T$$ 
    where the second inequality uses \Cref{eq:1-2alphaT} and $C \subseteq J$ and the third inequality uses $|C \cap W_{\sf card}| \leq 1$.
    Notice that $a^+(S)-a(S) = \frac{1}{14} \cdot a(S)$. We obtain 
    \begin{align*}
        z(C) &\geq a^+(C \cap W^*) \\
        &= a(C \cap W^*) + a^+(C \cap W^*)-a(C \cap W^*) \\
        &\geq a(C \cap W^*) + a^+(S)-a(S) \\
        &\geq (1-\alpha)T + \frac{1}{14} \cdot \frac{2}{5} \alpha T \\
        &= y_p.
    \end{align*}
    where the second inequality follows from $S \subseteq C \cap W$ and $a^+(i)-a(i) \geq 0$ for all $i \in I$, and the third inequality uses \Cref{obs:vertex-exists}.
    \item  Suppose that $|C \cap W_{\sf card}| \geq 3$.
    Then, $$z(C) \geq a^+(C \cap W_{\sf card}) = 3 \alpha T > y_p.$$
    \item  Suppose that $|C \cap W_{\sf card}| = 2$.
    If there exists an item $i \in C \cap W_{\sf rest}$ with $a(i) \geq \frac{2}{5} \alpha T$, then
    $$z(C) \geq a^+(C \cap W^*) = a^+(C \cap W_{\sf card}) + a^+(C \cap W_{\sf rest}) \geq 2 \alpha T + a^+(i) = \frac{17}{7} \alpha T =y_p.$$
    Therefore, we may assume that for all $i \in C \cap W_{\sf rest}$, $a(i) \leq \frac{2}{5} \alpha T$.
    It follows that $a^+(C \cap W_{\sf rest}) - a(C \cap W_{\sf rest}) = \frac{1}{14} \cdot a(C \cap W_{\sf rest})$. 
    Since $a(C \cap W_{\sf rest}) =a(C \cap W^*)-a(C \cap W_{\sf card}) \geq (1-\alpha)T-2\alpha T = \frac{2}{5} \alpha T$, we obtain
    \begin{align*}
        z(C) &\geq a^+(C \cap W^*) \\
        &= a(C \cap W^*) + a^+(C \cap W^*)-a(C \cap W^*) \\
        &\geq a(C \cap W^*) +a^+(C \cap W_{\sf rest}) - a(C \cap W_{\sf rest}) \\
        &\geq (1-\alpha)T + \frac{1}{14} \cdot \frac{2}{5} \alpha T \\
        &= y_p
    \end{align*}
    where the second inequality follows from $C \cap W_{\sf rest} \subseteq C \cap W$ and $a^+(i)-a(i) \geq 0$ for all $i \in I$, and the third inequality uses \Cref{obs:vertex-exists}.
    \end{itemize}

    \item Suppose that there exists some $v \subseteq C$ with $v^p \in V^*$.
    Since $v$ is an edge in $B$ and $z'$ is a fractional vertex cover of $B$, we have $z(v) = z'(v) \cdot \frac{5}{4}\alpha T$. 
    Therefore, we obtain
\begin{align*}
        z(C) &= z(C \cap W^*) + z(C \cap K) \\
        &\geq a(C \cap W^*) + z(v) \\
        &\geq (1-2\alpha)T + \frac{5}{4}\alpha T \\
        &=(1-\frac{3}{4}\alpha)T\\
        &>y_p.
\end{align*}
where the first inequality uses $z(C \cap W^*) \geq a^+(C \cap W^*) \geq a(C \cap W^*)$ and $v \subseteq C \cap K$ and the second inequality uses \Cref{eq:1-2alphaT} and the fact that $z(v) = z'(v) \cdot \frac{5}{4}\alpha T$. 
   \end{itemize} 

We have shown that in all cases, $z(C) \geq y_p$. Since $y,z \geq 0$, all constraints of \eqref{lp:dclp} are satisfied. This concludes the proof.   
\end{proof}

We now use the above lemma to obtain a second upper bound on $|Q|-|F_Q|$.

\begin{restatable}{lemma}{lemLpApplicationTwo} \label{lem:lp-application-2} Suppose that  $\alpha = \frac{5}{17}$. Then,
    $|Q|-|F_Q| \leq \frac{15}{7M}\alpha T \cdot n_{\sf cheap} + \frac{7}{3M}\alpha T \cdot n_{\sf medium} + \frac{5}{2M} \alpha T \cdot (n_{\sf costly}+n_{\sf many}+\eta(H^*))$.
\end{restatable}
\begin{proof}
    We have $y(P)=y(Q)=|Q| \cdot M$ and $$z(I) = |F_Q| \cdot M +a^+(W^*) + z'(K) \cdot \frac{5}{4} \alpha T \leq M \cdot |F_Q|+ a^+(W^*) + \frac{5}{2} \alpha T \cdot \eta(H^*),$$ where the inequality uses \Cref{th:frac-vc} and \Cref{obs:line-graph}. Since $(y,z)$ is a feasible solution to \eqref{lp:dclp} by \Cref{lem:dual-feasible-2}, \Cref{obs:lp-duality} implies 
    $$|Q| \cdot M \leq |F_Q| \cdot M +a^+(W^*) + \frac{5}{2} \eta(H^*)$$
    Rearranging, we derive
    \begin{align*}
        (|Q|-|F_Q|) \cdot M &\leq a^+(W^*) + \frac{5}{2}\alpha T \cdot \eta(H^*)\\
        &=a^+(W_{\sf cheap}) + a^+(W_{\sf medium}) + a^+(W_{\sf costly} \cup W_{\sf many}) + \frac{5}{2}\alpha T \cdot \eta(H^*)\\
        &\leq \frac{15}{7} \alpha T \cdot n_{\sf cheap} + \frac{7}{3} \alpha T \cdot n_{\sf medium} + \frac{5}{2} \alpha T \cdot (n_{\sf costly}+n_{\sf many}+\eta(H^*))
    \end{align*}
    where the equality uses \Cref{eq:W-decomp} and the second inequality uses \Cref{lem:a+bound}.
    Dividing both sides by $M$ gives the result.
\end{proof}

Taking a convex combination of \Cref{lem:lp-application-1} and \Cref{lem:lp-application-2} allows us to prove the following lemma.
\begin{restatable}{lemma}{lemConcludingTwo}\label{lem:concluding-2}
    For $\alpha = \frac{5}{17}$, we have $\eta(H|_Q) \geq |Q|$.
\end{restatable}
\begin{proof}
Take the convex combination of the inequalities from \Cref{lem:lp-application-1} and \Cref{lem:lp-application-2} with factors $\frac{1}{35}$ and $\frac{34}{35}$, respectively.
This gives, for our choice of $\alpha$,
\begin{equation} \label{eq:final}
    |Q|-|F_Q| \leq n_{\sf cheap} + n_{\sf medium}+n_{\sf costly} + n_{\sf many} + \eta(H^*)
\end{equation}

We can conclude as follows.
\begin{align*}
    \eta(H|_Q) &\geq \explosions{\rho} + \eta(H^*) \\
    &\geq  |F_Q| + \explosions{\sigma} + \eta(H^*)\\
    &= |F_Q| + n_{\sf cheap} + n_{\sf medium} + n_{\sf costly} + n_{\sf cheap} + \eta(H^*)\\
    &\geq |F_Q| + |Q|-|F_Q|\\
    &=|Q|
\end{align*}
where the first inequality uses \Cref{prop:explosions-eta}, the second inequality uses \Cref{obs:fat-thin}, the first equality uses \Cref{eq:n-decomp} and the third inequality uses \Cref{eq:final}.
\end{proof}

Since we chose $Q \subseteq P$ arbitrarily, \Cref{th:main} follows from \Cref{lem:concluding-2} and \Cref{th:hall-general}.
\paragraph{Open problems.} We conclude with some open problems.
Naturally, it is desirable to investigate by how much the integrality gap of \eqref{lp:clp} can be improved further using the approach based on \Cref{th:hall-general}.
Another direction, already mentioned by \cite{haxell2025improved}, is to work towards making the proof constructive.
Finally, it would be very interesting to use this method for more general, non-additive valuation functions. For instance, it is open whether \eqref{lp:clp} has constant integrality gap when players have submodular valuations.
\\\\\textbf{AI disclosure. } No generative AI was used for obtaining proofs or writing the article.
\bibliographystyle{splncs04}
\bibliography{bib}

\end{document}